\documentclass[reprint,amssymb,amsmath,aip,cha]{revtex4-1}
\usepackage[caption = false]{subfig}
\usepackage{color}
\usepackage{epsfig}
\usepackage{rotating}
\usepackage{ifpdf}
\usepackage{bm}
\usepackage[colorlinks=true,linkcolor=blue]{hyperref}%
\expandafter\ifx\csname package@font\endcsname\relax\else
 \expandafter\expandafter
 \expandafter\usepackage
 \expandafter\expandafter
 \expandafter{\csname package@font\endcsname}%
\fi
\usepackage{cleveref}
\begin{document}
\title{Rotational properties of annulus dusty plasma in a strong magnetic field}
\author{Mangilal Choudhary}
\email{jaiijichoudhary@gmail.com} 
\affiliation{Institute of Advanced Research, The University for Innovation, Koba, Gandhinagar, 382426, India}
\author{Roman Bergert}
\author{Sandra Moritz}
\author{Slobodan Mitic}
\author{Markus H. Thoma}
\affiliation{I. Physikalisches Institut, Justus--Liebig Universität Giessen, Henrich--Buff--Ring 16, D 35392 Giessen, Germany}
\begin{abstract} 
The collective dynamics of annulus dusty plasma formed between a co-centric conducting (non-conducting) disk and ring configuration is studied in a strongly magnetized radio-frequency (rf) discharge. A superconducting electromagnet is used to introduce a homogeneous magnetic field to the dusty plasma medium. In absence of the magnetic field, dust grains exhibit thermal motion around their equilibrium position. The dust grains start to rotate in anticlockwise direction with increasing magnetic field (B $>$ 0.02 T), and the constant value of the angular frequency at various strengths of magnetic field confirms the rigid body rotation. The angular frequency of dust grains linearly increases up to a threshold magnetic field (B $>$ 0.6 T) and after that its value remains nearly constant in a certain range of magnetic field. Further increase in magnetic field (B $>$ 1 T) lowers the angular frequency. Low value of angular frequency is expected by reducing the width of annulus dusty plasma or the input rf power. The azimuthal ion drag force due to the magnetic field is assumed to be the energy source which drives the rotational motion. The resultant radial electric field in the presence of magnetic field determines the direction of rotation. The variation of floating (plasma) potential across the annular region at given magnetic field explains the rotational properties of the annulus dusty plasma in the presence of magnetic field.
\end{abstract}
\maketitle
\section{Introduction}
Dusty plasma, which is an admixture of electrons, ions and negatively charged micron sized solid particles exhibits solid- as well as liquid-like characteristics, depending on coulombic interactions among charged particles \cite{morfilldusty1}. Due to such unique features, dusty plasma is considered as part of soft condensed matter similar to granular material and colloidal suspension \cite{granularmedium1,granularmedium2,colloidalsuspension1,colloidalsuspension2}. The negatively charged dust grains have a large inertia; therefore, dusty plasma responses to a very low frequency field which allows to study solid state phenomena, fluid dynamics, turbulences, etc. at kinetic level using optical observation technique \cite{morfilldusty1,bonitzdusty2,shukladusty3,merlinodusty4}.\par
 In laboratory experiments, dusty plasma dynamics depends on the forces acting on the dust grains \cite{barnesdustforces,mangilalthesis,nitterdustlevitation}. The dust charging process as well as forces acting on the dust grain are associated to the ambient plasma background. Collective dynamics of dusty plasma is expected to change with altering the background plasma. It is known that input power and pressure are major external variables to alter plasma dynamics. Instead of these variables, an external magnetic field can also modify the  background plasma \cite{mangilalpsst}. Therefore, magnetic field is  considered as an external parameter (variable) to study the collective dynamics of dusty plasma. Hence, a wide spectrum of theoretical as well as experimental works have been  performed to study the role of an external magnetic field on dusty plasma \citep{melzerbookdusty}. \par
In the presence of an axial magnetic field, strongly coupled dust particles confined over a cathode in a DC discharge exhibit the rotational motion with a constant angular velocity in a plane and velocity shear in the vertical direction\cite{uchida2dflow}. The role of a longitudinal magnetic field to a small planar dust cluster confined in a striation head of the positive column of a DC discharge has been reported by many researchers \cite{vasilievdcrotationinb,karasevlongitudinalbrotation,dzlievarotationstratamagnetic,dyachkovrotationmagneticdc}. In such configuration, they observed the inversion of dust structure rotation above a threshold magnetic field at given discharge conditions. Konopka \textit{et al.} \cite{knopkamagneticrotation} investigated the role of a vertical magnetic field to a 2-dimensional dust crystal confined by a conducting ring on the sheath of a radiofrequency (rf) discharge and observed two types of rotations named as rigid-body rotation and sheared rotation. Cheung \textit{et al.} \cite{cheung2to12planarrotation} investigated the characteristics of a rotating dust cluster containing 1 up to 12 dust grains in a rf discharge under the influence of an external magnetic field. This work confirms the role of dust particles number in a dust cluster to determine the rotational characteristics in presence of a magnetic field. The role of gas pressure in determining the rotational properties of a dust cluster has been studied by Huang \textit{et al.} \cite{huangdustrotationrf}. They reported different rotational speeds for upper and lower dust layers as well as reversal of rotation direction at a threshold pressure.\par
Recently, Dzlieva \textit{et al.} \cite{dzlievarotationstrongb} investigated the influence of a strong magnetic field (B $\sim$ 1 T) to a dust structure in a DC discharge. They observed the inversion of rotation at lower magnetic field strength and a slow increase of the angular frequency after B $>$ 0.1 T. Karasev \textit{et al.} \cite{karasevstrongb} reported a rotating dust cluster and shell structure in the presence of a strong magnetic field. Apart from DC discharges, some interesting features of 3D dusty plasmas, as damping of dust-acoustic waves \cite{mangilalwavedamping} and counter-rotating dust torus \cite{mangilalvortex3d}  have been reported in recent experiments by Choudhary \textit{et al.} in rf discharge. Melzer \textit{et al.} \cite{melzerstrongbrotation} investigated the effect of a strong magnetic field (B $>$ 5 T) to a dust cluster confined by a ring shaped electrode in a rf discharge. They observed the increase of rotation frequency of the two-dimensional (2D) dust cluster with increasing magnetic field. The rotation of 2D dust clusters confined by a radial electric field either in DC discharge or rf discharge in the presence of a longitudinal magnetic field is a result of the $E \times B$ drift of ions in the azimuthal direction of a cylindrical geometry \cite{kawrotation,rotationinionflow,knopkamagneticrotation}.\par
In above mentioned studies, dust grains (in rf discharge) are confined by a radial electric field of an additional ring on the lower electrode or the ring-like structure of the lower electrode. Dust grains form a cluster due to the radial electric field of the ring edge and exhibit rotational motion in presence of an external magnetic field \cite{knopkamagneticrotation}. Few work has been performed to study the collective dynamics of an annulus dusty plasma in a magnetized rf discharge. Bandyopadhyay \textit{et al.} \cite{pintushearflowmagnetic} observed the sheared or opposite dusty plasma flow in the annular region of co-centric conducting rings in the presence of a magnetic field. They claimed an increase of angular frequency of opposite flowing dust grains with increasing magnetic field.
It is difficult to get answers to many open questions that arise after these first experimental observations. Is it possible to induce shear flow in an annulus dusty plasma at any discharge condition in the presence of a magnetic field? Does an annulus dusty plasma remain stable at a strong magnetic field? Is it possible to induce rigid rotational motion in an annulus dusty plasma in the presence of an external magnetic field? How does the angular frequency depend on the external magnetic field?. There are many other questions regarding the flow characteristics of the annulus dusty plasma at a strong magnetic field. To get the answers to some of these questions, experiments were performed in a strongly magnetized rf discharge where dust grains were confined between the co-centric (conducting or non-conducting) disk-and-ring-configuration.\par
Section~\ref{sec:exp_setup} deals with the detailed description of the experimental set-up and the plasma and dusty plasma production. The dynamics of annuls dusty plasma produced between the co-centric (conducting and non-conducting) disk-and-ring-configuration is discussed in Section~\ref{sec:dynamics}. The origin of rotational motion is discussed in Section~\ref{sec:discussion}. A brief summary of the work along with concluding remarks is provided in Section~\ref{sec:summary}.        
\section{Experimental setup}  \label{sec:exp_setup}
\begin{figure*}
 \centering
\subfloat{{\includegraphics[scale=0.85]{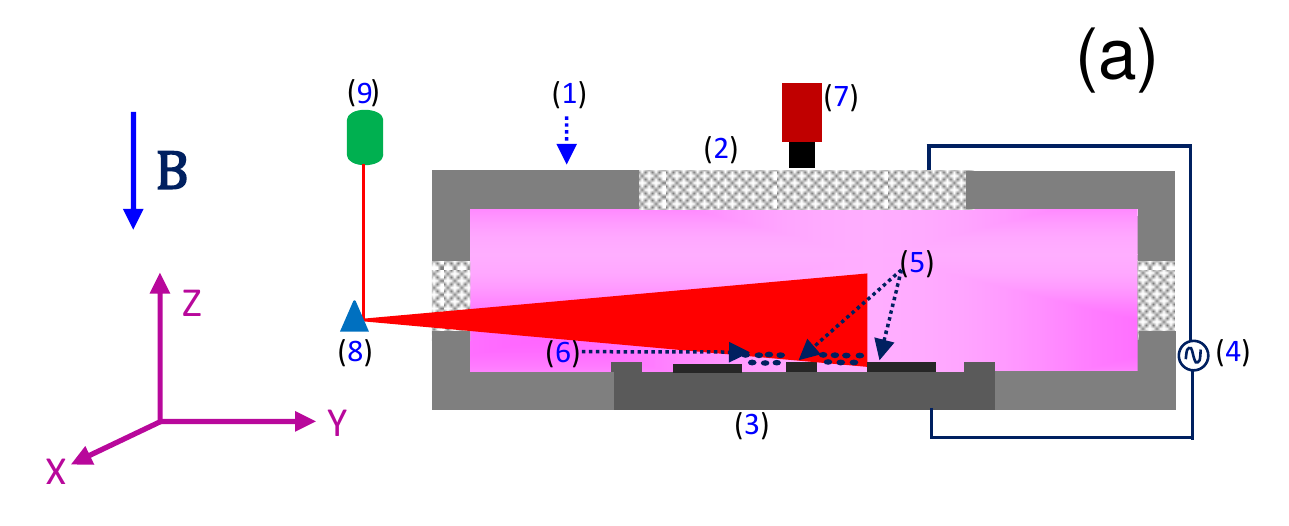}}}%
\qquad
\subfloat{{\includegraphics[scale=0.60]{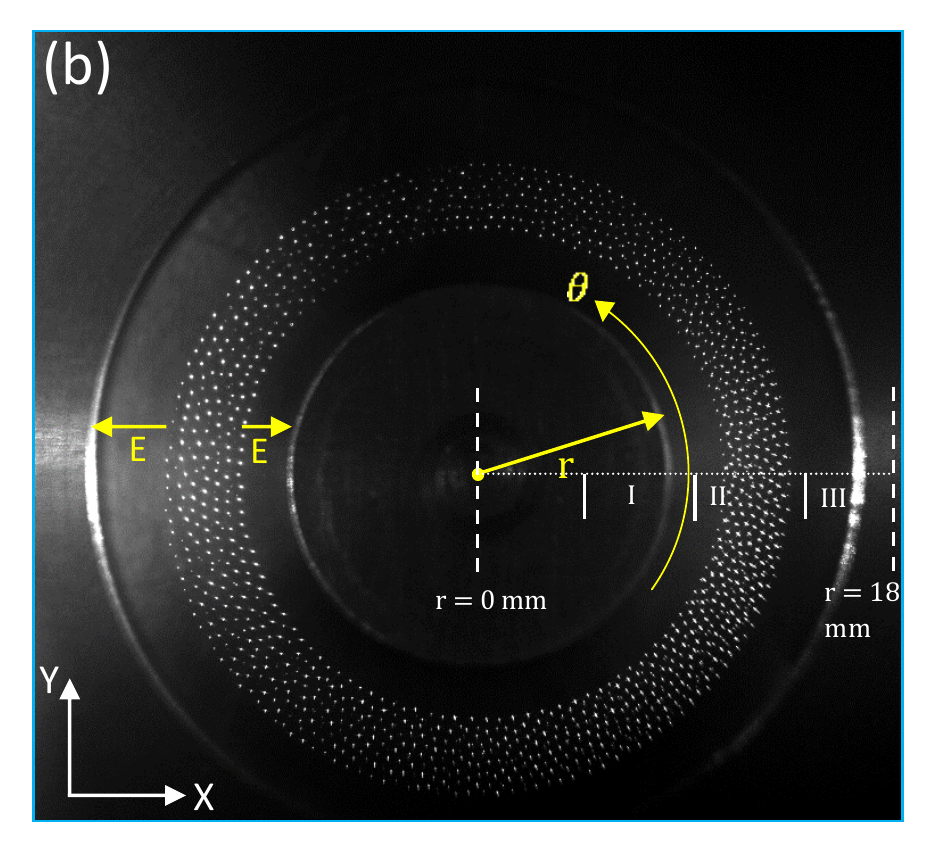}}}
 \qquad
 \caption{\label{fig:fig1}(a) Schematic diagram of the experimental setup to study the magnetized dusty plasma (1) with vacuum chamber, (2) upper ITO-coated transparent electrode, (3) lower aluminum electrode, (4) RF power generator, (5) aluminum (Teflon) disk and ring, (6) confined dust particles, (7) CMOS camera, (8) mirror, and (9) red laser with a cylindrical lens. The blue error indicates the direction of external magnetic field. (b) The experimental view of the annulus dusty plasma formed between the co-centric aluminum disk ($D_{disk}$ = 15 mm) and ring ($D_{ring}^{in}$ = 30 mm). In this study we used both Cartesian and Cylindrical Coordinates to understand the rotational motion. The experimental region is divided into three different regions; disk edge region (I), central or dusty plasma region (II) and ring edge region (III). The direction of electric field in the annular region is represented by a yellow arrow.}
 \end{figure*}

The present set of experiments is carried out in a vacuum chamber, which is placed at the center of a superconducting electromagnet ($B_{max}$ $\sim$ 4 T) to introduce a homogeneous external magnetic field to the dusty plasma. The details of the magnetized dusty plasma device available at Justus-Liebig University Giessen is provided in Ref.\cite{mangilalpsst}. A schematic diagram of the experimental setup is presented in Fig.~\ref{fig:fig1}(a). Before starting the experiments, a disk (conducting or non-conducting) of diameter $D_{disk}$ = 15 mm and a ring with an inner diameter of $D_{ring}^{in}$ 30 mm, outer diameter of $D_{ring}^{out}$ = 50 mm and thickness of 2 mm are placed on the lower electrode. The centres of the disk and the ring coincide with the center of the lower electrode to make an annular region. Aluminium and Teflon are used as conducting and non-conducting materials, respectively. After placing the co-centric disk and ring on the lower electrode, the vacuum chamber is evacuated to base pressure p $<$ $10^{-2}$ Pa using a pumping system consisting of a rotary and a turbo molecular pump. The experiments are performed with argon gas and the pressure inside the chamber is controlled by using a mass flow controller (MFC) and gate valve controller.  At a given gas pressure, plasma is ignited between an aluminium electrode of 65 mm diameter (lower electrode) and an indium tin oxide-coated ((ITO-coated) electrode of 65 mm diameter (upper electrode) using a 13.56 MHz rf generator with a matching network. Between the electrodes there is a gap of 30 mm. A dust dispenser, which is installed at one of the side ports of the vacuum chamber, is used for injecting the Melamine Formaldehyde (MF) particles with a diameter of $D$ $\simeq$ 6.28 $\mu$m into the plasma volume. The MF particles acquire negative charges and are confined in the annular region of the co-centric disk and ring. To investigate the flow dynamics or rotational properties of the annulus dusty plasma in presence of the magnetic field, a red laser sheet and a CMOS camera are used to illuminate the particles and to capture the scattered light coming from dust grains, respectively. The CMOS camera is installed to observe the dust dynamics through the transparent upper electrode in the horizontal plane (X--Y plane) at a frame rate of 60 fps and with a resolution of 2048$\times$2048 pixels. The stored images are later analyzed with help of ImageJ \cite{imagejsoftware} software and MATLAB based open-access code, called openPIV \cite{piv}. A full view of the annulus dusty plasma formed between the co-centric aluminium disk and ring in the horizontal (X--Y) plane is shown in Fig.~\ref{fig:fig1}(b). \par
\begin{figure*}
\centering
 \includegraphics[scale=0.48]{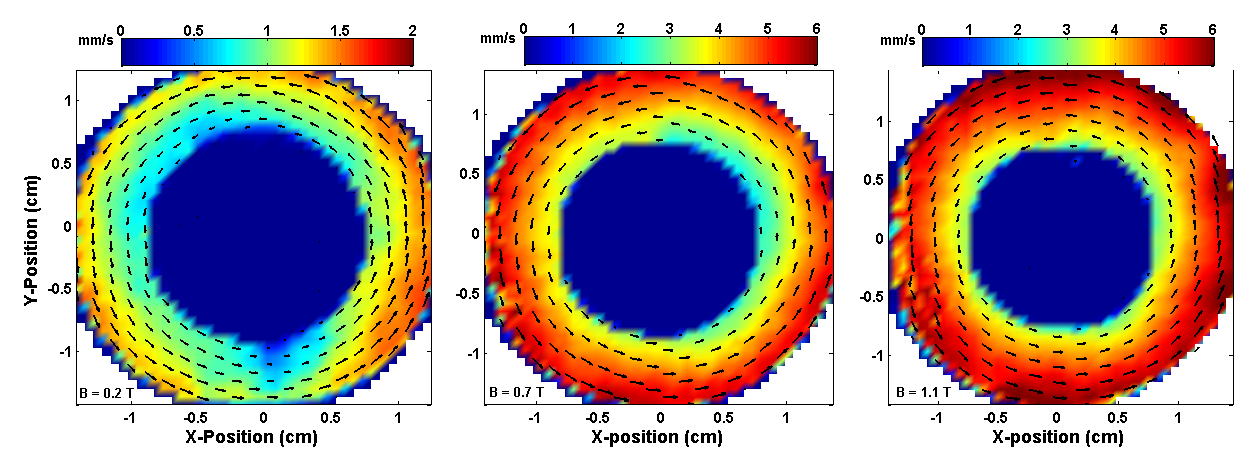}
\caption{\label{fig:fig2}PIV images of the rotational motion of dust grains in a 2D (X--Y) plane at various strengths of magnetic fields. Dust grains are confined between the Teflon disk ($D_{disk}$ = 15 mm) and ring ($D_{ring}^{in}$ = 30 mm) at  electrode voltages $V_{up}$ = 60 V and $V_{down}$ = 50 V and argon pressure, p = 30 Pa. The PIV images are corresponding to B = 0.2 T, B = 0.7 T and B = 1.1 T, respectively. The central blue color region includes the disk, as well as a dust free region (see Fig.~\ref{fig:fig1}(b)) which can be identified using the scale of plots.}
\end{figure*}
\section{Annulus dusty plasma dynamics in the presence of a magnetic field}  \label{sec:dynamics}
It has been reported experimentally that 3-dimensional dusty plasma exhibits the vortex motion instead of azimuthal rotation in the presence of a strong magnetic field \cite{mangilalvortex3d}. However, a two-dimensional dust cluster always rotates in the direction of the magnetic field-induced ion flow \cite{knopkamagneticrotation}\cite{melzerstrongbrotation}. To investigate the dynamics of a nearly 2-dimensional annulus dusty plasma (confined between the co-centric disk and ring) in presence of a strong magnetic field, two types of combinations of disk and ring are used. In the first set of experiments, a combination of co-centric non-conducting (Teflon) disk and ring is used to create the annulus dusty plasma. In another set of experiments, annulus dusty plasma is produced between the co-centric conducting (Aluminum) disk and ring at given discharge conditions. The detailed study of the annulus dusty plasma in presence of a strong magnetic field is presented in the subsections \ref{sec:teflon_potential_well} and \ref{sec:alumiium_potential_well}.
\begin{figure*}
 \centering
\subfloat{{\includegraphics[scale=0.33]{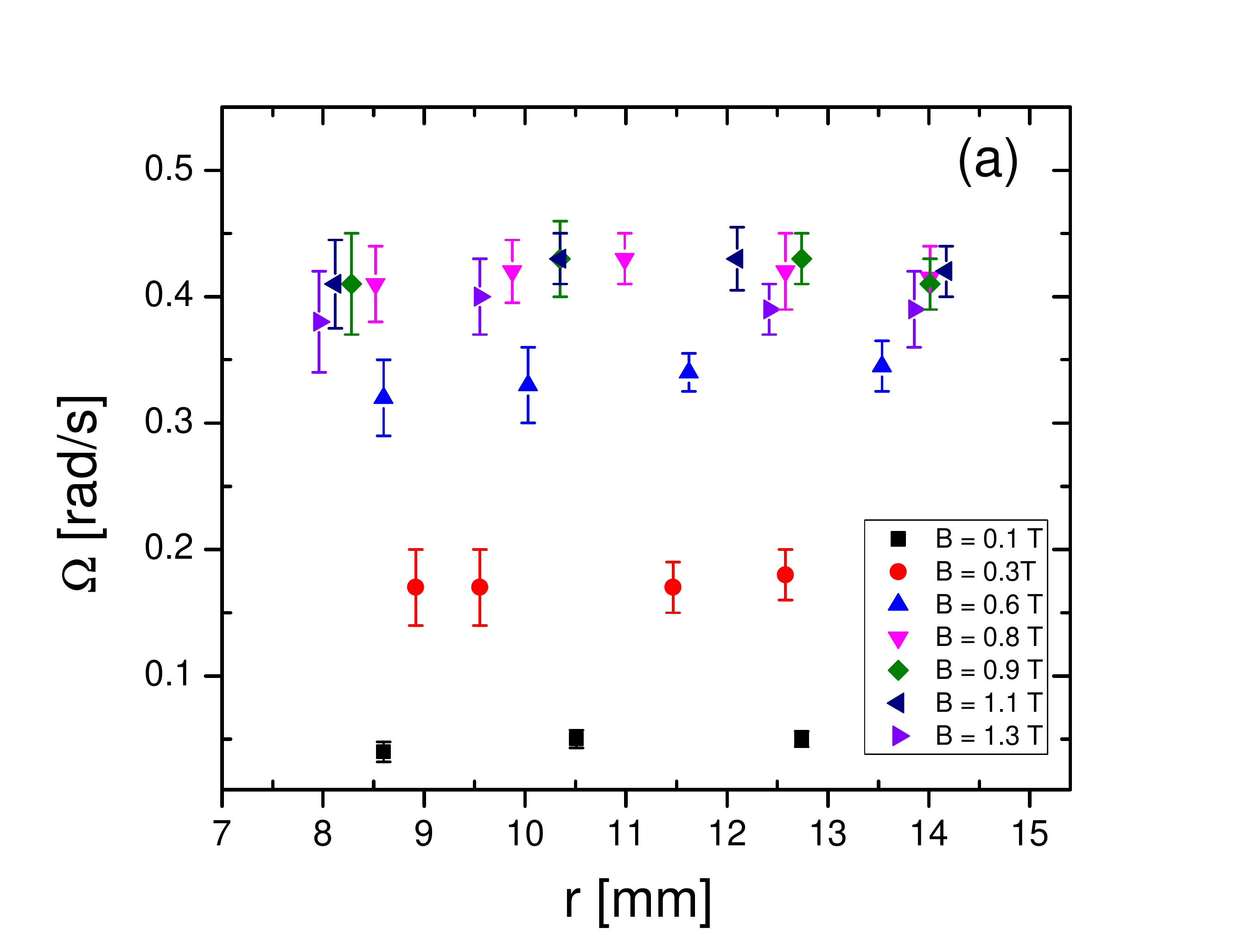}}}%
\subfloat{{\includegraphics[scale=0.33]{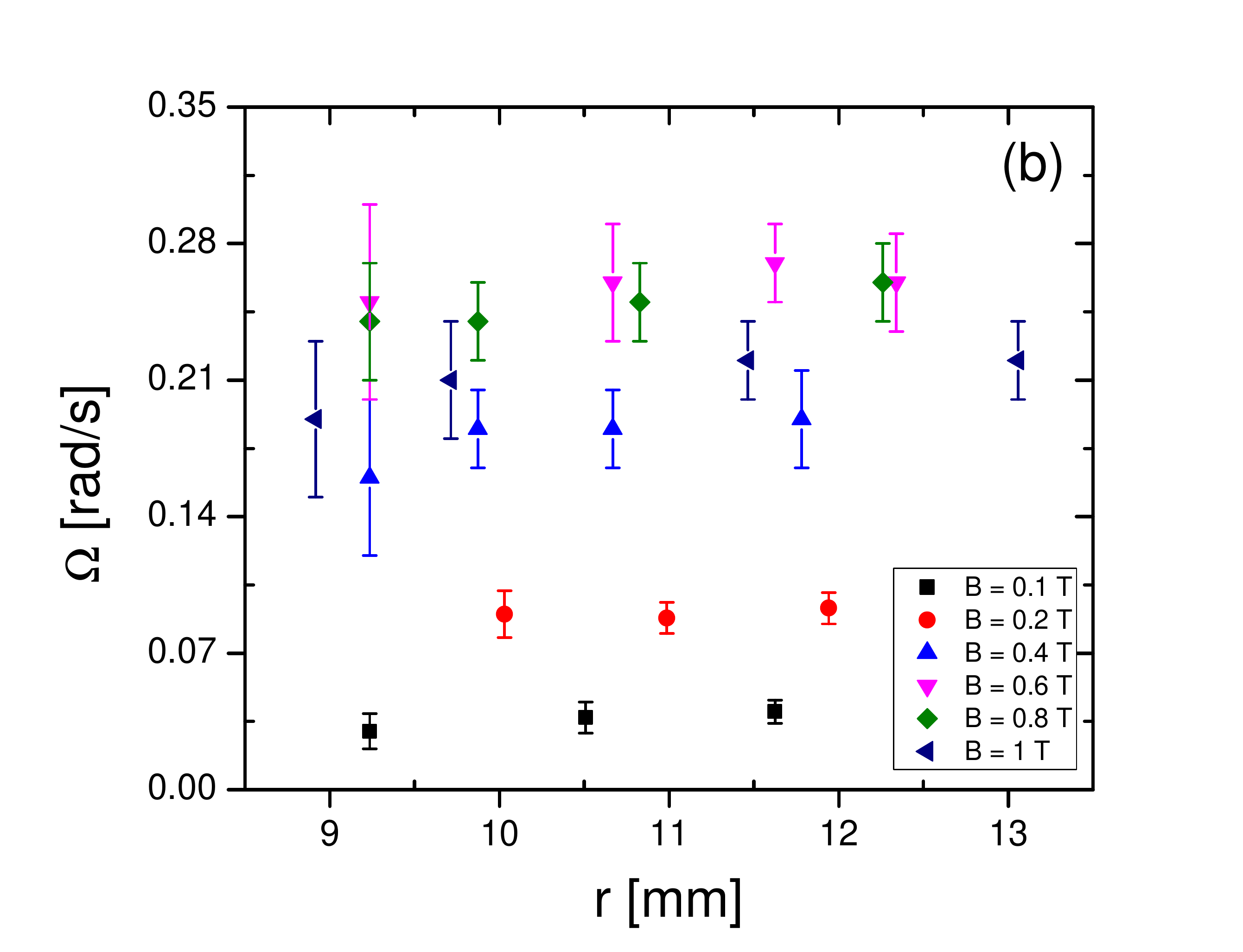}}}
\qquad
\subfloat{{\includegraphics[scale=0.33]{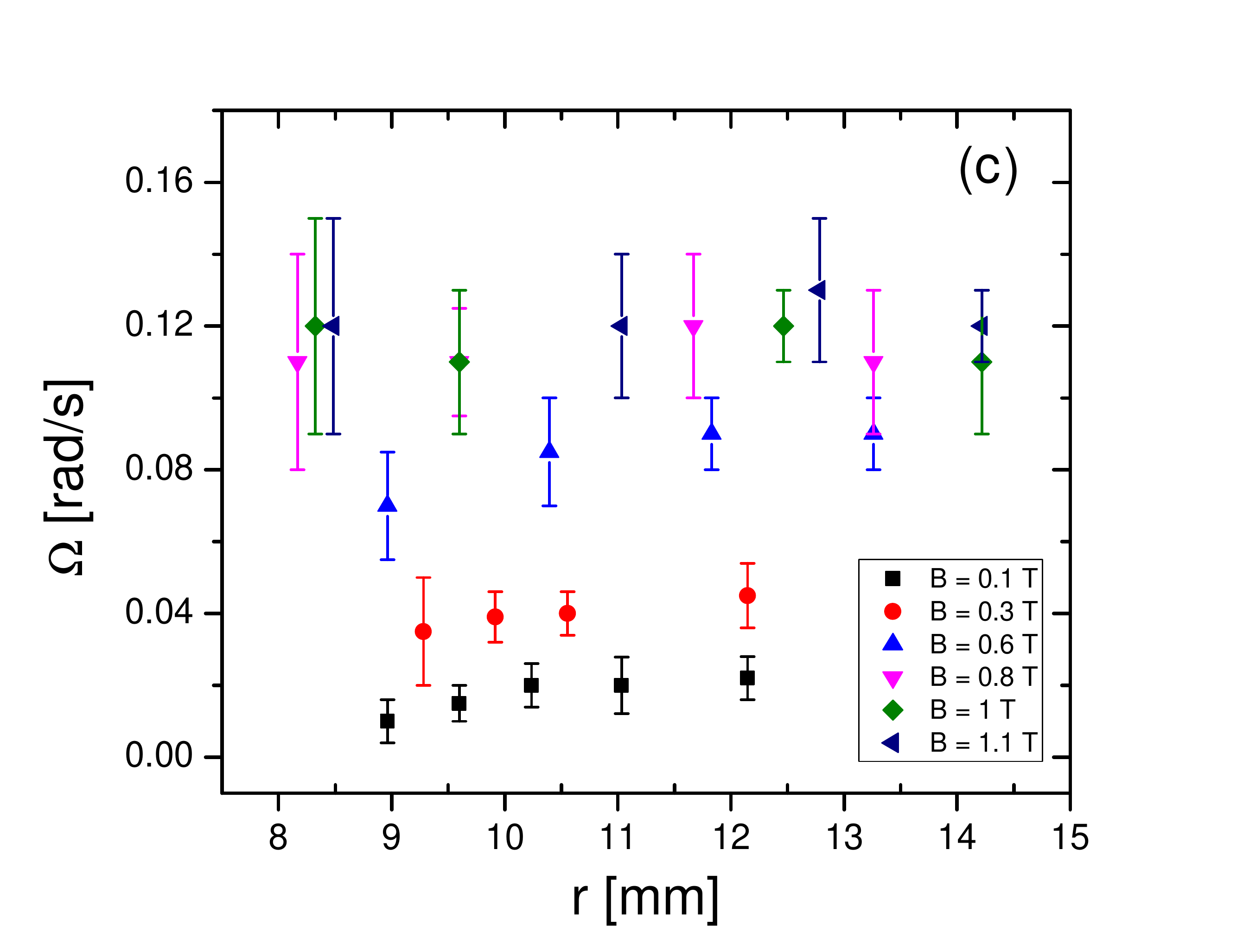}}}%
\qquad
\caption{\label{fig:fig3} The radial variation of angular frequency ($\Omega$) of rotating particles at various strengths of the magnetic field. (a) Teflon disk ($D_{disk}$ = 15 mm) and ring ($D_{ring}^{in}$ = 30 mm), $V_{up}$ = 60 V and $V_{down}$ = 50 V,  p = 30 Pa and annulus width of dusty plasma $\sim$ 5 mm (at B = 0 T). (b) Teflon disk ($D_{disk}$ = 15 mm) and ring ($D_{ring}^{in}$ = 30 mm), $V_{up}$ = 60 V and $V_{down}$ = 50 V,  p = 30 Pa and annulus width of dusty plasma $\sim$ 2.5 mm. (c) Teflon disk ($D_{disk}$ = 15 mm) and ring ($D_{ring}^{in}$ = 30 mm), $V_{up}$ = 50 V and $V_{down}$ = 50 V, p = 30 Pa and annulus width of dusty plasma $\sim$ 4 mm.}
\end{figure*}
\subsection{Dusty plasma confined between the co-centric non-conducting disk and ring}  \label{sec:teflon_potential_well}
The potential well created between the disk and the ring is a result of modification of the potential distribution in rf sheath of the lower powered electrode. After injecting the dust grains into plasma, they acquire negative charges and get confined in the potential well. Gas pressure p and electrodes voltage (or rf power) are adjusted to make a 2-dimensional (3 to 5 layers in vertical plane) annulus dusty plasma at B = 0 T. The annulus width of the dust grain medium can be changed by injecting less or more particles into the plasma for a given discharge condition. The dust density ($n_d$) is expected to increase with increasing the annulus width of dusty plasma. In the first set of experiments, dusty plasma is produced at p = 30 Pa and  voltages of the upper electrode, $V_{up}$ = 60 V and of the lower electrode, $V_{down}$ = 50 V. The annulus width of dusty plasma at B = 0 T is estimated to be around 5 mm.  As the magnetic field is applied to the dusty plasma, at first the confining potential well gets modified at low magnetic field strengths, resulting in a slight change in position of dust grains. Therefore, the tracking of the rotational motion at low magnetic field (B $<$ 0.02 T) is difficult. The anticlockwise rotational motion of annuals dust grain medium is predicted above B $>$ 0.02 T.  It is noticed that the annulus width of dusty plasma medium starts to increase above a threshold value of magnetic field strength of B $>$ 0.6 T. AT B = 1.3 T, an estimated width of dusty plasma is $\sim$ 6.5 mm. This variation in annulus width of dusty plasma is expected due to the modification in the confining potential well at strong magnetic field. The dust density at this annulus width comes out to be $\sim$ 3-5 $\times$ $10^4 cm^{-3}$. For a detailed study of rotational motion (e.g. direction, angular frequency) at various strengths of magnetic field, the still images are further analyzed. \par
 \begin{figure}
\centering
 \includegraphics[scale=0.33]{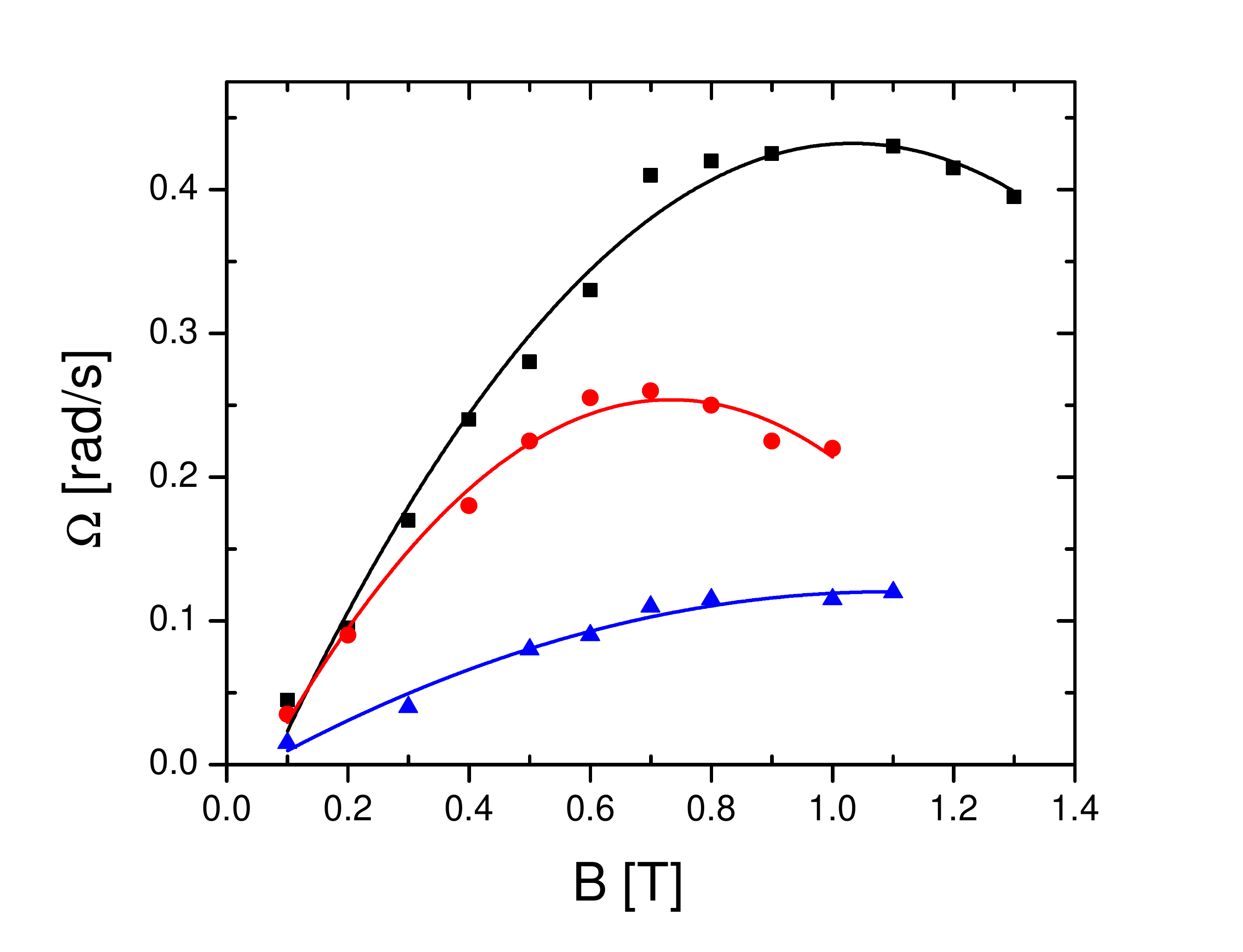}
\caption{\label{fig:fig4}The angular frequency variation against magnetic field corresponds to the discharge conditions of Fig.~\ref{fig:fig3}  {\color{black}$\blacksquare$}  $V_{up}$ = 60 V, $V_{down}$ = 50 V and annulus width of dusty plasma $\sim$ 5 mm, {\color{red}$\bullet$}  $V_{up}$ = 60 V, $V_{down}$ = 50 V, and annulus width $\sim$ 2.5 mm, {\color{blue}$\blacktriangle$} $V_{up}$ = 50 V and $V_{down}$ = 50 V and annulus width $\sim$ 4 mm.}
\end{figure}
In Fig~\ref{fig:fig2}, PIV images of rotating dust grains in horizontal plane (X--Y plane) at different strengths of magnetic field are displayed. The PIV images are constructed using an adaptive 2-pass algorithm (a 64$\times$ 64, 50\% overlap followed by a 32$\times$ 32, 50\% overlap analysis). The contour maps (Fig.~\ref{fig:fig2}) of the average magnitude of velocity are constructed after averaging the velocity vectors of consecutive 50 frames. In the color map of the PIV images, the direction of velocity vectors represents the direction of rotating particles in the annular region. The magnitude of the average velocity of rotating particles is represented by color bars. It is clear from Fig.~\ref{fig:fig2} that particles rotate in anticlockwise direction at given magnetic field. Magnitude of the azimuthal velocity ($v_\phi$) of rotating particles increases from the disk edge region (r $\sim$ 8 mm) to the ring edge region (r $\sim$ 14.5 mm), which indicates a velocity gradient. For getting the angular frequency of rotating particles, $\omega = {v_\phi}/{r} $, the PIV images at different magnetic fields are further analysed. An average angular frequency ($\Omega$) of the rotating particles between the co-centric disk and ring is shown in Fig.~\ref{fig:fig3}(a).                                                                                                                                                                                                                                                                                                                                                                                                                                                                                                                                                                                                                                                                                                                                                                                                                                                                                                                                                                                                                                                                                                                                                                                                                                                                                                                                                                                                                                                                                          
 The angular frequency of rotating particles is found to be nearly constant within the errors at different strengths of magnetic field, B = 0.1 T to 1.3 T. The nearly constant value of $\Omega$ represents the rigid body rotation of annulus dusty plasma in the presence of a magnetic field.\par 
Does the rotational properties of annulus dusty plasma depend on its width? To see the effect of annulus width of the dust grain medium on the rotational motion, a second set of experiments is carried out for dusty plasma with a lower annulus width $\sim$ 2.5 mm for the same disk-and-ring-configuration. The discharge conditions (p = 30 Pa, $V_{up}$ = 60 V and $V_{down}$ = 50 V) are kept similar to the earlier experiments (Fig.~\ref{fig:fig3}(a)). In this case, we estimate the $n_d$ $\sim$ 1-2 $\times$ $10^4 cm^{-3}$ at B = 0 T. The average angular frequency of rotating particles from disk edge (r $\sim$ 9 mm) to ring edge (r $\sim$ 13 mm) at different magnetic field strengths is presented in Fig.~\ref{fig:fig3}(b). The dusty plasma width increment from $\sim$ 2.5 mm to $\sim$ 4 mm after increasing the magnetic field above a threshold value (B $>$ 0.6 T) is also noticed. The variation of $\Omega$ is observed to be nearly constant at given magnetic field strengths (B = 0.1 T to 1 T). The constant value of $\Omega$ represents the rigid body rotation of dust grain medium. 
In comparison of $\Omega$ at given B for large annulus width (Fig.~\ref{fig:fig3}(a)) and small annulus width (Fig.~\ref{fig:fig3}(b)) dusty plasma conforms the increase of $\Omega$ with increasing the annulus width of dusty plasma (or dust density) at same discharge conditions. However, the flow (or rotational) characteristics (e.g. direction, rigid rotation) of different annulus widths dusty plasmas remain same at given magnetic field. \par
\begin{figure*}
\centering
 \includegraphics[scale=0.50]{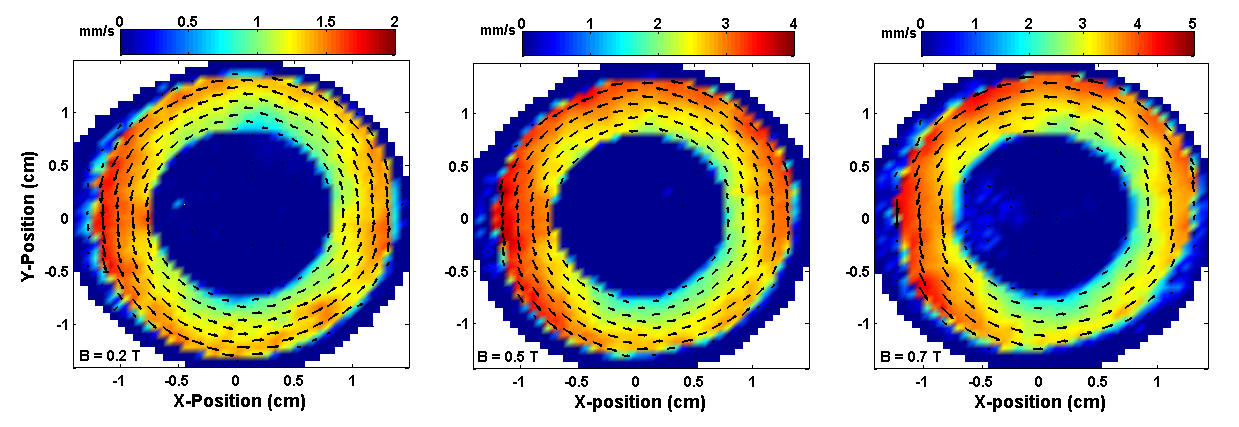}
\caption{\label{fig:fig5}PIV images of the rotational motion of dust grains in the X--Y plane at various strengths of magnetic field. Particles are confined between the aluminum disk ($D_{disk}$ = 15 mm) and ring ($D_{ring}^{in}$ = 30 mm)  at  electrode voltages $V_{up}$ = 55 V and $V_{down}$ = 55 V and argon pressure, p = 30 Pa. The inner blue color region includes the disk as well a dust free region (see Fig.~\ref{fig:fig1}(b)).}
\end{figure*}
\begin{figure*}
\centering
 \includegraphics[scale=0.50]{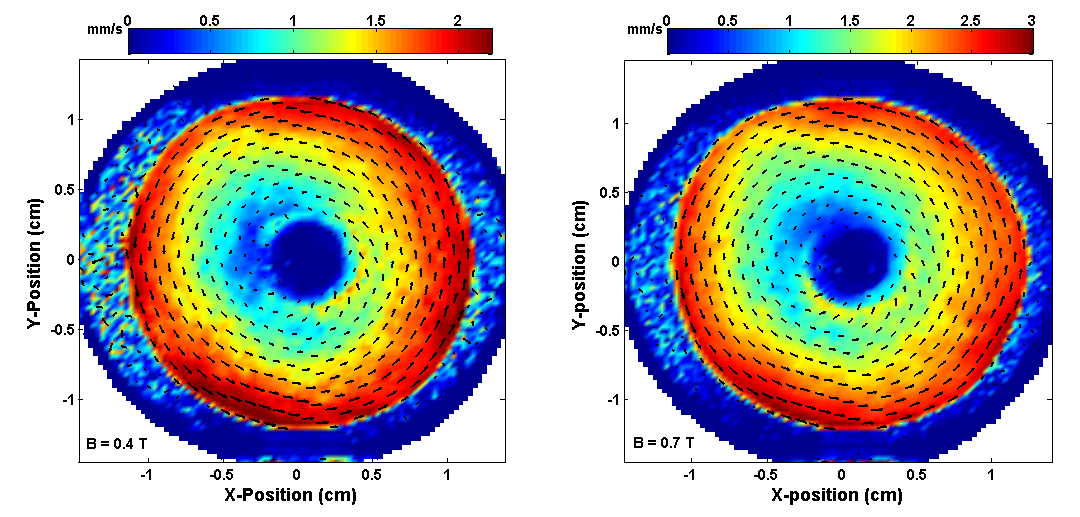}
\caption{\label{fig:fig6}PIV images of the rotational motion of dust grains in the X--Y plane at various strengths of magnetic field. Dust grains are confined between the aluminum disk ($D_{disk}$ = 5 mm) and ring ($D_{ring}^{in}$ = 30 mm) at electrodes voltage $V_{up}$ = 55 V and $V_{down}$ = 55 V and argon pressure, p = 30 Pa. The inner blue color region includes the disk as well a dust free region that can be separated using the plots scale.}
\end{figure*}
Another set of experiments is performed to observe the effect of electrode voltage (rf input power) on the rotational properties of annulus dusty plasma in presence of a magnetic field. Fig.~\ref{fig:fig3}(c) shows the variation of angular frequency of particles at pressure p = 30 Pa and peak-to-peak voltages of 50 V in the presence of a magnetic field. It should be noted that the confining potential well between disk and ring gets modified if $V_{up}$ is lowered. Therefore, the annulus width of dusty plasma ($\sim$ 4 mm) is always smaller than that of high values of $V_{up}$. $\Omega$ has a lower value near the disk edge (r $\sim$ 9 mm to 10 mm) in the low magnetic field regime (B $<$ 0.8 T). However, nearly constant $\Omega$ is observed within an error range of $<$ 15 \%. At a distance far away from the disk edge (r $>$ 10 mm), $\Omega$ remains almost constant for all values of magnetic field strength. Thus, annulus dusty plasma exhibits rigid rotational motion in anticlockwise direction for a given value of magnetic field strength (B $<$ 1.1 T). Higher values of $\Omega$ are expected at large peak-to-peak electrode voltages (or input rf power) in the presence of a magnetic field as seen in Fig.~\ref{fig:fig3}(a) and Fig.~\ref{fig:fig3}(b). \par
 It is also important to plot the average angular frequency against the strength of the external magnetic field. In Fig.~\ref{fig:fig4}, the average values of averaged $\Omega$ (Fig.~\ref{fig:fig3}(a) to Fig.~\ref{fig:fig3}(c)) is plotted against magnetic field strength. In this figure, we see a linear increase of angular frequency up to a threshold magnetic field strength B $>$ 0.6 T and after that it remains nearly unchanged or saturated in a certain range of magnetic field strength (0.6 T $<$ B $<$ 1 T). Further increase in magnetic field strength (B $>$ 1 T) slightly lowers the $\Omega$-value. The rate of increase of $\Omega$  below a threshold magnetic field strongly depends on the electrodes' peak-to-peak voltage for a given pressure. The saturation value of $\Omega$ shifts to a lower value if the dust width is decreased or input power is lowered.\\\\
\begin{figure*}
 \centering
\subfloat{{\includegraphics[scale=0.33]{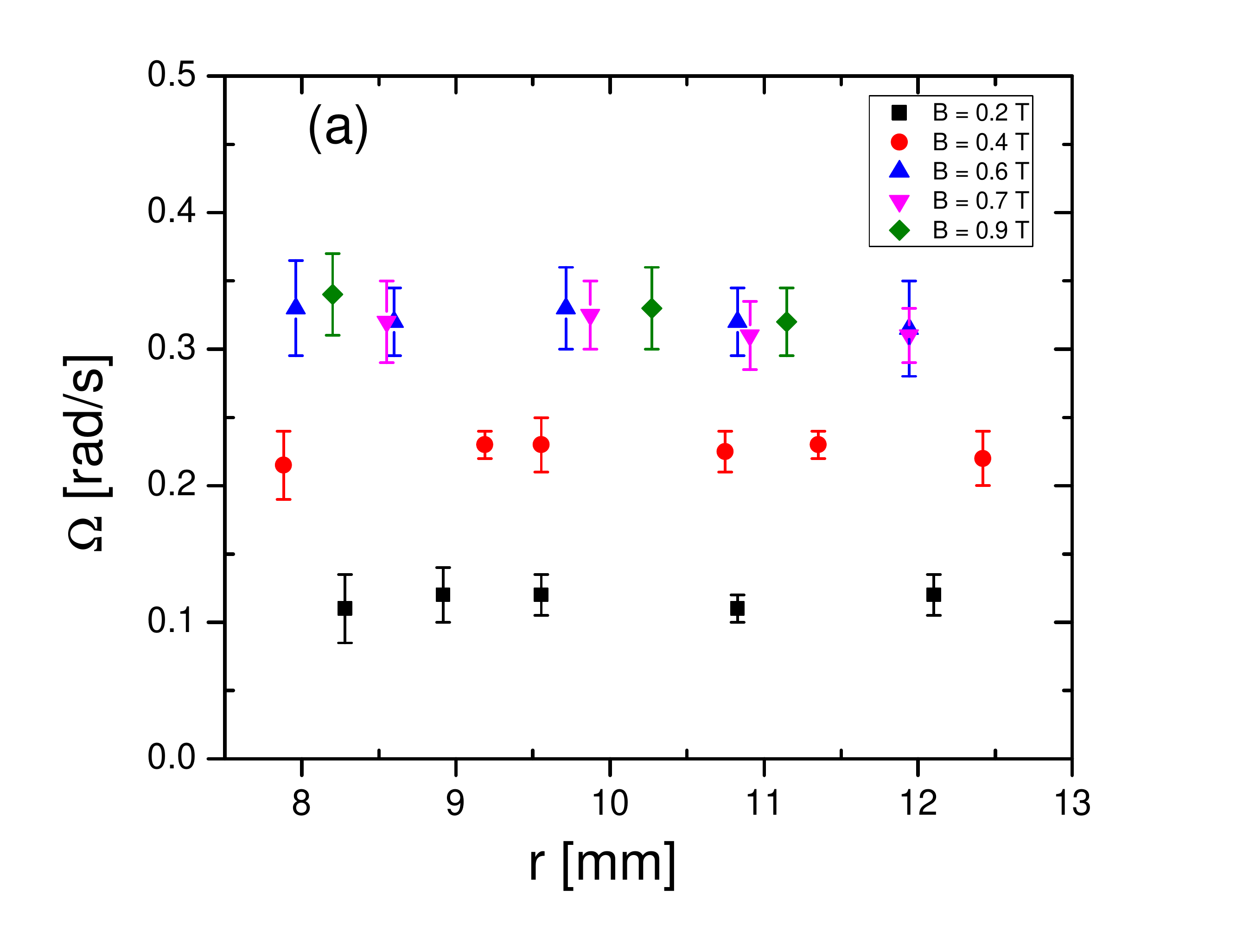}}}%
\subfloat{{\includegraphics[scale=0.33]{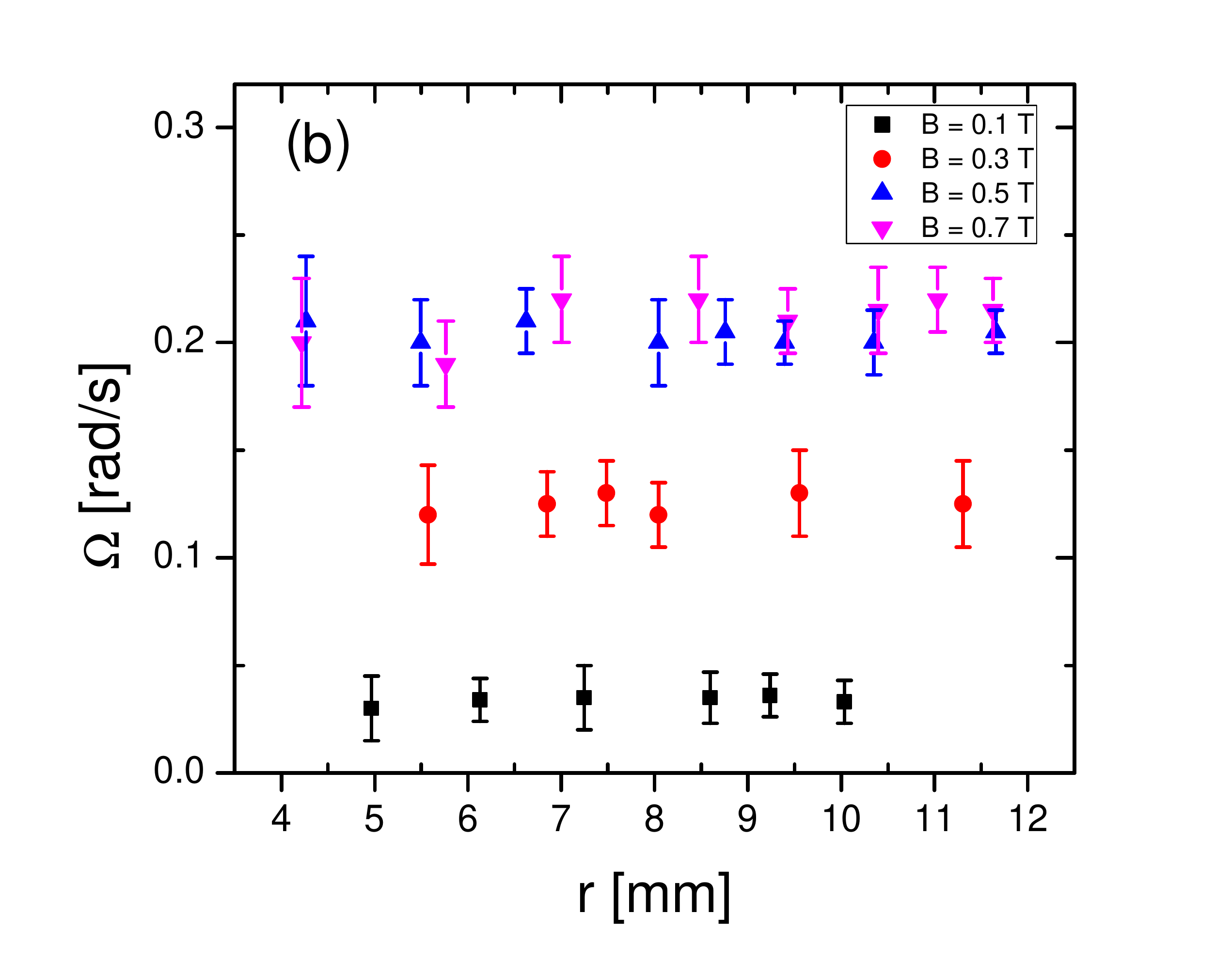}}}
 \qquad
 \subfloat{{\includegraphics[scale=0.33]{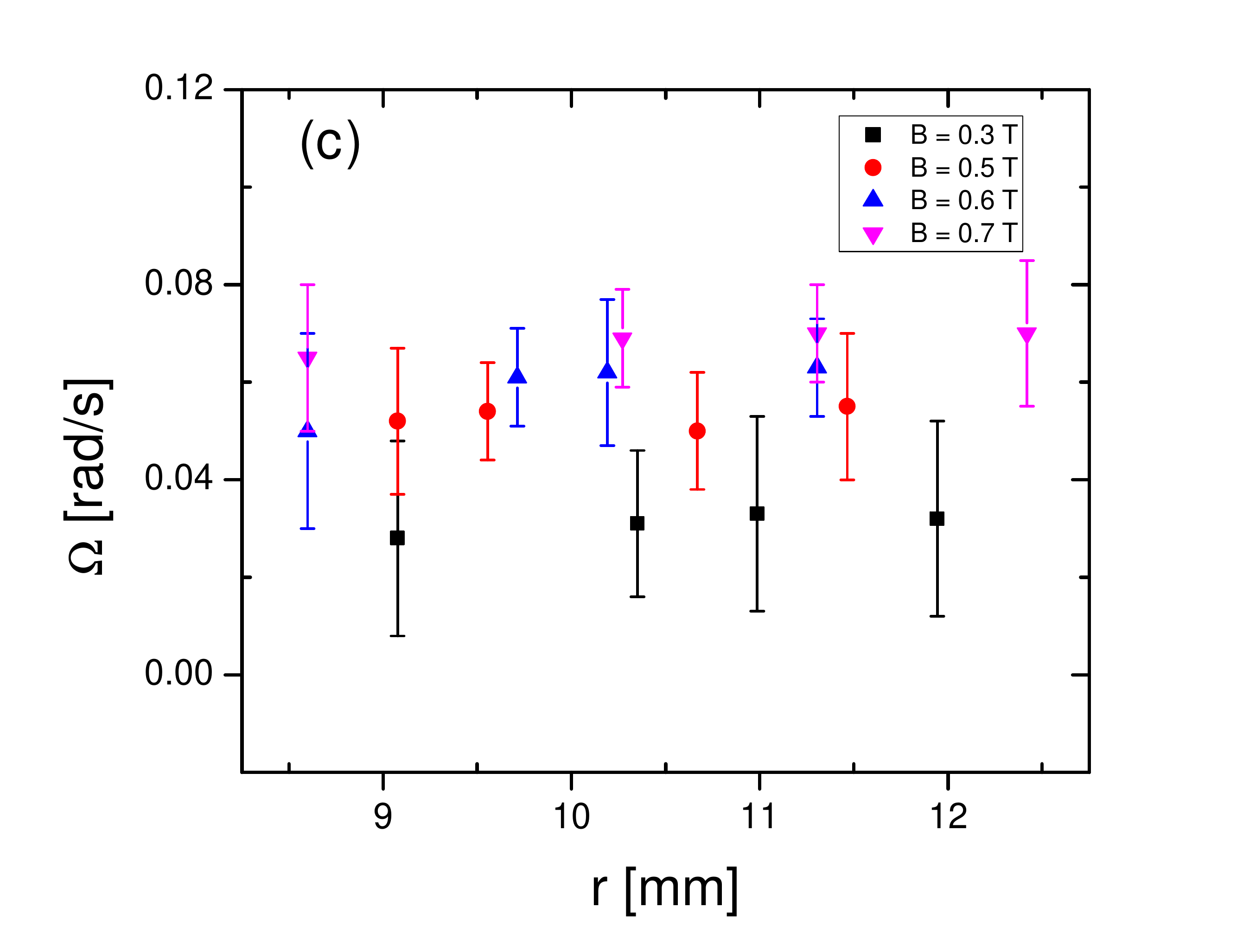}}}
\caption{\label{fig:fig7} The radial variation of angular frequency ($\Omega$) of rotating particles at various strengths of magnetic field is shown. (a) Aluminum disk ($D_{disk}$ = 15 mm) and ring ($D_{ring}^{in}$ = 30 mm), $V_{up}$ = 55 V and $V_{down}$ = 55 V,  p = 30 Pa and annulus width of dusty plasma $\sim$ 4 mm at B = 0 T. (b) Teflon disk ($D_{disk}$ = 5 mm) and ring ($D_{ring}^{in}$ = 30 mm), $V_{up}$ = 55 V and $V_{down}$ = 55 V,  p = 30 Pa and annulus width $\sim$ 9 mm at B = 0 T. (c) Teflon disk ($D_{disk}$ = 15 mm) and ring ($D_{ring}^{in}$ = 30 mm), $V_{up}$ = 45 V and $V_{down}$ = 45 V, and p = 30 Pa.}
 \end{figure*}
\subsection{Dusty plasma confined between the co-centric conducting disk and ring}  \label{sec:alumiium_potential_well}
 One of our previous studies reports the influence of ring materials (conducting and non-conducting) on the depth of the confining potential well in the presence of a magnetic field \cite{mangilalvortex3d}. This is probably due to the difference of surface potentials of the materials in an rf discharge. A dip potential well is expected between the conducting co-centric disk and ring. At this point, it is not easy to describe the dust grain dynamics in an annular region of the conducting disk and ring in the presence of a strong magnetic field. Therefore, a set of experiments is carried out using the aluminium disk-and-ring-configuration in the place of the Teflon configuration (\ref{sec:teflon_potential_well}). It is observed in experiments that plasma between ring and disk is unstable at high rf power (or large voltage on electrodes) and high magnetic field (B $>$ 0.4 T). Therefore, a study is conducted at low peak-to-peak electrodes voltage of 55 V ($V_{up}$ = 55 V and $V_{down}$ = 55 V). At this discharge conditions, two separate experiments with an aluminium disk of diameter ($D_{disk}$) 15 mm and 5 mm are performed. The aim of using a small diameter (5 mm) disk is to demonstrate the role of the magnetic field on the flow characteristics of a greater annulus width of the dusty plasma. The annulus width of dusty plasma for a 15 mm diameter disk is  $\sim$ 4 mm and for a 5 mm disk is $\sim$ 9 mm at B = 0 T. For getting more details about rotational properties of dusty plasma, PIV analysis of still images is performed as discussed in Sec.~\ref{sec:teflon_potential_well}.\par
Fig.~\ref{fig:fig5} shows the PIV images of rotating particles between the co-centric disk and ring ($D_{disk}$ = 15 mm, $D_{ring}^{in}$ = 30 mm) at various strengths of magnetic field. The velocity distribution confirms the radial velocity gradient from the disk edge region (r $\sim$ 8 mm) to ring edge region (r $\sim$ 12 mm). The direction of rotation is anticlockwise for all values of magnetic field strength (B $<$ 1 T) at P = 30 Pa and peak-to-peak electrodes voltage of 55 V. The velocity distribution of rotating particles between the co-centric disk and ring ($D_{disk}$ = 5 mm, $D_{ring}^{in}$ = 30 mm) is presented in Fig.~\ref{fig:fig6}. We observe the radial velocity gradient as well as uni-directional (anticlockwise) rotational motion similar to Fig.~\ref{fig:fig5} at various strengths of magnetic field. \par 
The variation of angular frequency of dust grains at different magnetic field strengths (correspond to Fig.~\ref{fig:fig5} and Fig.~\ref{fig:fig6}) is depicted in Fig.~\ref{fig:fig7}(a) and Fig.~\ref{fig:fig7}(b), respectively. The nearly constant values of $\Omega$ from disk edge to ring edge region demonstrate the rigid body rotation of dust grain medium for a given magnetic field. To observe the effect of the electrodes voltage on annulus dusty plasma, the experiments are performed at low peak-to-peak electrodes voltage of 45 V ($V_{up}$ = 45 V and $V_{down}$ = 45 V)  and p = 30 Pa in the presence of a magnetic field. The variation of $\Omega$ from disk edge to ring edge region at various strengths of magnetic field is shown in Fig.~\ref{fig:fig7}(c). At lower electrodes voltage (or low rf power), dust grains also rotate in anticlockwise direction with a constant angular frequency. A lower value of $\Omega$  is observed in case of low rf power or low $V_{pp}$. The direction of rotation of annulus dusty plasma is found to be anticlockwise at various sets of pressure (p = 30 to 50 Pa) and electrodes voltages (40 to 75 V) for such disk-and-ring-configurations. It should also be noted that the reversal of rotation is not observed at different discharge conditions in the presence of a magnetic field. \par
\begin{figure}
\centering
 \includegraphics[scale=0.32]{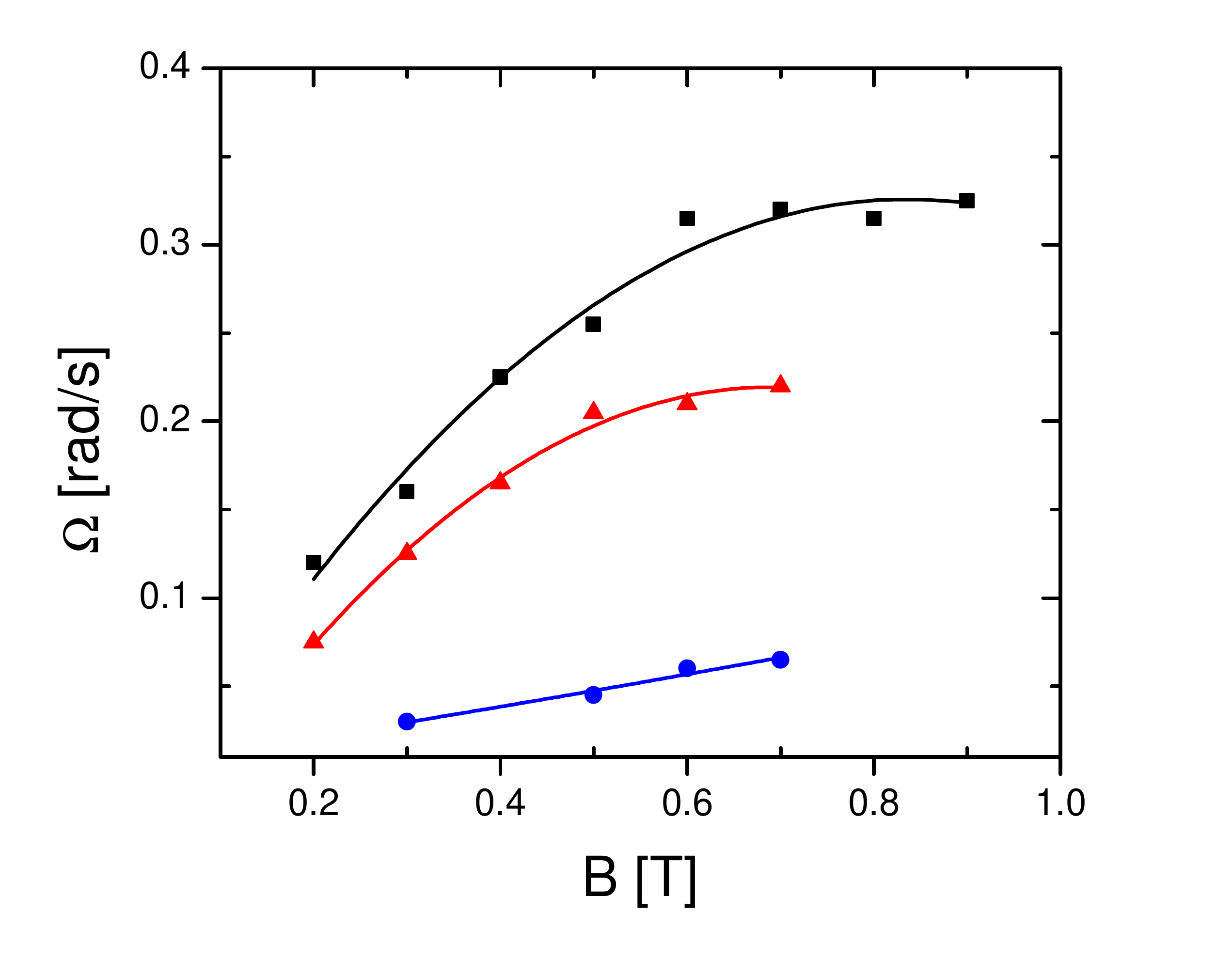}
\caption{\label{fig:fig8} The angular frequency plots against magnetic field correspond to discharge conditions of Fig.~\ref{fig:fig7} {\color{black}$\blacksquare$}  $V_{up}$ = 55 V, $V_{down}$ = 50 V and annulus width of dusty plasma $\sim$ 4 mm, {\color{red}$\bullet$}  $V_{up}$ = 55 V, $V_{down}$ = 55 V, and annulus width $\sim$ 9 mm, {\color{blue}$\blacktriangle$} $V_{up}$ = 45 V and $V_{down}$ = 45 V and annulus width $\sim$ 3 mm.}
\end{figure}
 The average value of $\Omega$ at a given magnetic field strength (Fig.~\ref{fig:fig7}) is used to get the plots of $\Omega$ against magnetic field strength. Fig.~\ref{fig:fig8} represents the angular frequency variation of rotating particles with magnetic field. It is observed that $\Omega$ increases almost linearly with increasing magnetic field up to a threshold value of magnetic field strength (B $>$ 0.6 T). Above the threshold magnetic field strength, a slight change in $\Omega$ with increasing magnetic field is noticed. The change rate of $\Omega$ below the threshold magnetic field depends on the rf power (or plasma density). The threshold value of magnetic field strength is found to be almost the same for different annular widths dusty plasmas at same discharge conditions but it shifts to higher or lower values with changing the electrodes voltage ($V_{up}$ and $V_{down}$).
\section{Discussion} \label{sec:discussion}
 In an rf discharge, massive dust particles are suspended in the strong electric field of the sheath region to balance the gravitational force and simultaneously confined by a radial electric field created by an additional ring on the lower powered electrode \cite{nitterdustlevitation,mangilalthesis}. In this study, dust grains are confined in a potential well created by a co-centric conducting (non-conducting) disk-and-ring-configurations. The grains are confined in the annular region of a disk and ring, where the radial electric fields are in opposite directions. The radial electric field points inward and outward due to the disk and ring, respectively (see Fig.~\ref{fig:fig1}(b)). Before turning on the magnet (B = 0 T), dust grains exhibit the thermal motion around their equilibrium position. At low magnetic field strength (B$<$ 0.02 T), first the confining potential well gets modified and with increasing the magnetic field (B $>$ 0.02 T) dust grains start to rotate in anticlockwise direction . \par
 In unmagnetized plasma, ions flow ($\vec{v_i}$ = $\mu_i \vec{E_r}$) radially in the direction of electric field ($\vec{E_r}$). In the presence of a magnetic field, the radial ion flow gets deviated due to the Lorentz force (q ($\vec{v_i} \times \vec{B}$)). Due to the Lorentz or $\vec{E_r} \times \vec{B}$ force on ions, they also have a velocity component in the azimuthal direction. The azimuthal drifted ions drag the dust grains in the direction of flow and set them into rotational motion in the presence of a magnetic field \cite{kawrotation,knopkamagneticrotation,2dclusterrotation}. The direction of rotation of dust grain medium is determined by $\vec{E} \times \vec{B}$ drift of ions. Since the radial electric fields are in opposite directions due to the disk and ring, an opposite or shear dust flow (clockwise due to the disk and anticlockwise due the ring) is expected in the presence of a magnetic field. However, only anticlockwise rotational motion of annulus dusty plasma is observed at various strengths of magnetic field at given discharge condition. We observe the rotational inversion with changing the direction of magnetic field in our experiments. It confirms the role of azimuthal ion drag or $\vec{E} \times \vec{B}$ drifted motion of ions to rotate the annulus dust grain medium in presence of the magnetic field. \par
It is known that the radial electric field strength can be estimated by measuring the plasma potential ($V_p$) across the radial distance from disk edge region (I) to ring edge region (III). Since the annular region is $\sim$ 8 mm, the use of an emissive probe to measure plasma potential is a challenging task at strong magnetic field strength. It is difficult to measure the small variation in $V_p$ across this radial distance. It is also known that $V_p$ can be estimated by using the measured value of the floating potential ($V_f$) for a given value of $T_e$ \cite{probemerlino}. In the present experiments, $T_e$ variation across the radial distance (r $\sim$ 6 mm to 16 mm) is expected to be negligible \cite{mangilal3dvortexb}. Therefore, the radial variation of $V_f$ is assumed to be the $V_p$ variation at a given magnetic field strength. A cylindrical probe of length $\sim$ 2 mm and radius 0.125 mm is used to measure radial variation of $V_f$ \cite{mangilalpsst} from disk edge region (r $\sim$ 4mm) to ring edge region (r $\sim$ 18 mm).\par
 It should be noted that plasma parameters are not strongly affected by the dust grains if Havnes parameter \cite{havnes} $P_h$ = $Z_d n_d/n_i$ has a lower value, i.e, $P_h << 1$. Here, $n_d$ is the dust density, $n_i$ is the ion density and $Z_d$ is the dust charge number. In such dusty plasma ($P_h < 1$), ambient plasma parameters without dust grains can be used to understand the dynamics of dust grain medium. In our experiments, we estimate $P_h <$ 1 for $n_d \sim$ 1 $\times 10^{3}$ to 5 $\times 10^{3}$ $cm^{-3}$, $n_i \sim$ 6 $\times 10^{8}$ $cm^{-3}$ to 9 $\times 10^{8}$ $cm^{-3}$, and $Z_d \sim$ 1 $\times 10^4$. Hence, $V_f$ measurements are carried out in ambient plasma (without dust grains) to understand the rotational motion of annulus dusty plasma.\\
 \begin{figure}
 \centering
\subfloat{{\includegraphics[scale=0.31]{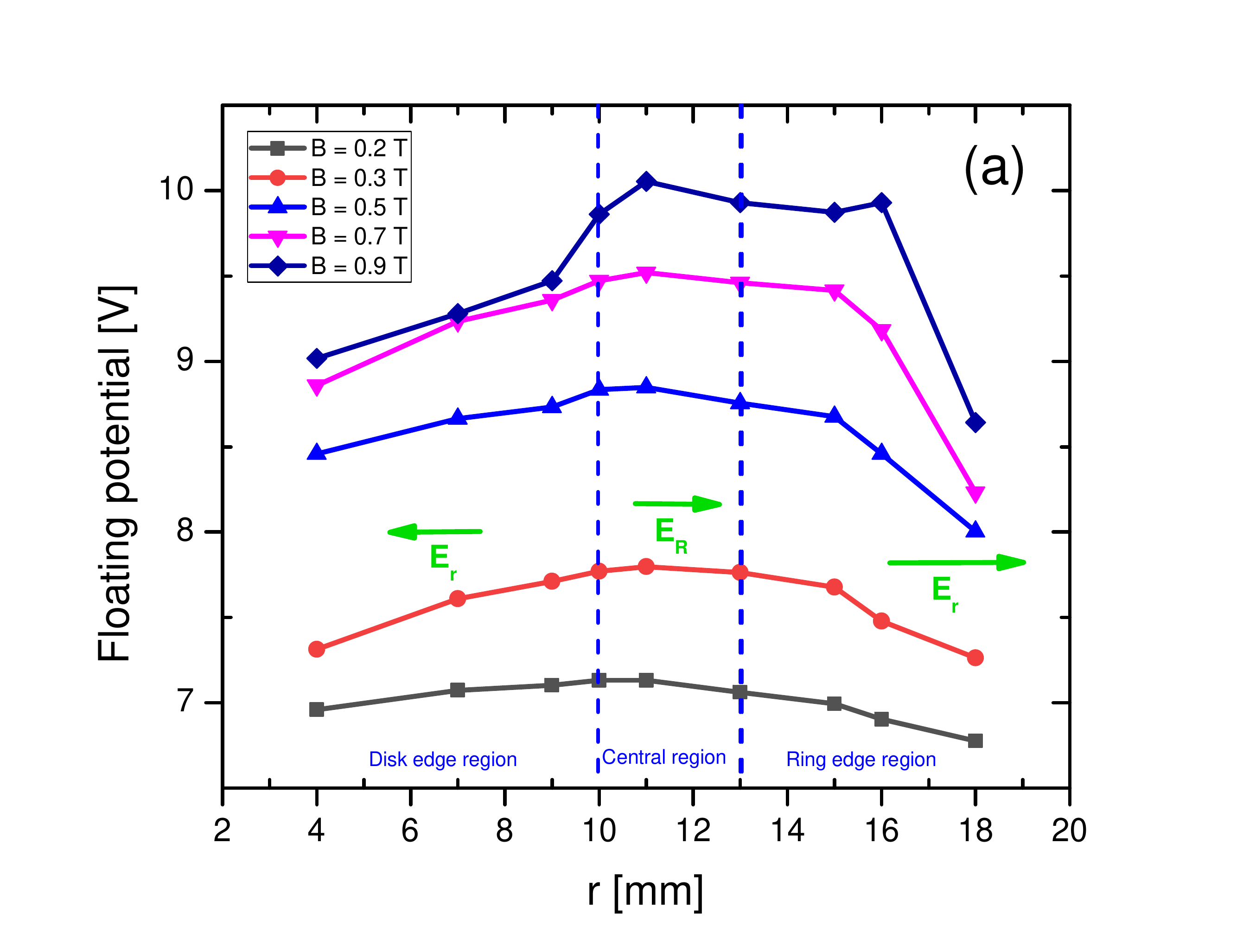}}}%
\qquad
\subfloat{{\includegraphics[scale=0.31]{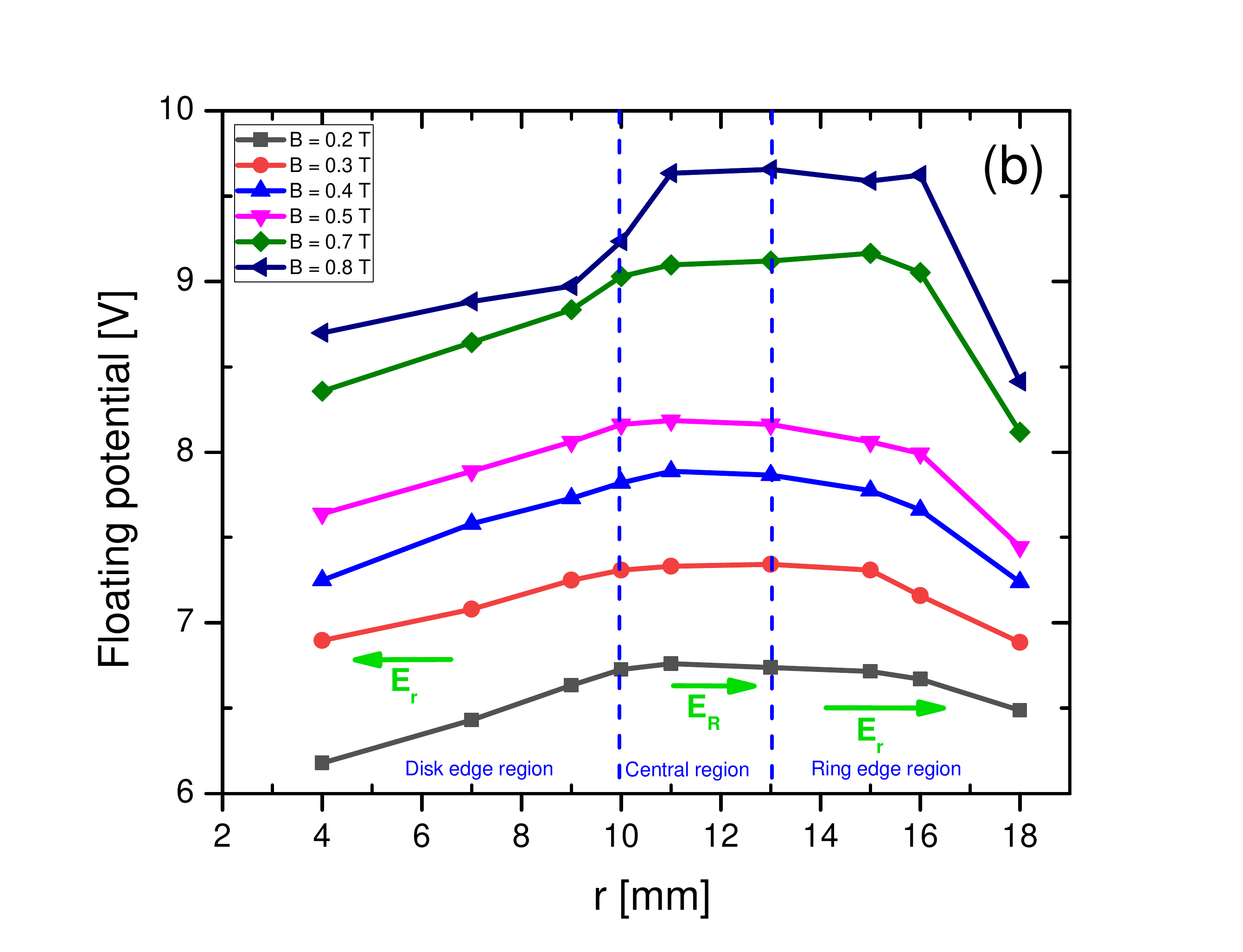}}}
 \qquad
\caption{\label{fig:fig9} Floating potential variation across the radial distance at various strengths of magnetic field. (a) Teflon disk ($D_{disk}$ = 15 mm) and ring ($D_{ring}^{in}$ = 30 mm),  $V_{up}$ = 60 V and $V_{down}$ = 50 V, and p = 30 Pa. (b) Aluminum disk ($D_{disk}$ = 15 mm) and ring ($D_{ring}^{in}$ = 30 mm,  $V_{up}$ = 55 V and $V_{down}$ = 55 V, and p = 30 Pa. The green arrows represent the direction of radial electric field (E).}
 \end{figure}
Fig.\ref{fig:fig9} shows the radial variation of $V_f$ at Z $\sim$ 0.7 cm for (a) non-conducting (Teflon) and (b) conducting (aluminium) disk and ring ($D_{disk}$ = 15 mm, $D_{ring}^{in}$ = 30 mm) configurations at given discharge conditions in the presence of a magnetic field. For a clearer understanding of $V_f$ variation, the plots are divided into three regions as presented in Fig.~\ref{fig:fig1}(b). The dust grains are confined in the central region. Fig.~\ref{fig:fig8} represents clearly that $V_f$ has a larger gradient in the ring edge region (I) than in the disk edge region (III) for a given magnetic field. The gradient in $V_f$ defines the strength of the radial electric field in that region. Therefore, a stronger radial electric field is expected in the ring edge region (III) than in the disk edge region (I) at a given magnetic field. The resultant electric field ($E_R$) in the central region (II) is the superposition of both opposite radial electric fields. Therefore, the annulus dust cluster is expected to rotate in accordance to the direction of the resultant electric field ($\vec{E_R}$). Thus, the direction of rotation is observed as anticlockwise at various strengths of magnetic field. The strength of azimuthal ion drag force determines the magnitude of $\Omega$ at a given magnetic field and increases with increasing magnetic field. Above the threshold magnetic field strength (B $>$ 0.6 T), the $V_f$ gradient is found to be strong near the edge of the ring (see Fig.~\ref{fig:fig9}), resulting in a low resultant electric field in the central region. We also observe a slight expansion of dust cluster width (slightly weak coupling) above a threshold magnetic field strength. These two factors, resultant electric field and coupling strength among the dust grains, determine the magnitude of angular frequency (lower value) at strong magnetic field strength. \\
 \section{Summary} \label{sec:summary}
The rotational properties of an annulus dust grain medium confined in a potential well created between a co-centric conducting (or non-conducting) disk and ring are explored in a strongly magnetized rf discharge. A superconducting electromagnet is used to introduce a uniform magnetic field to the dusty plasma. The plasma is ignited between a lower aluminum electrode and an upper ITO-coated transparent electrode using a 13.56 MHz rf generator with the matching network for a given argon pressure. The dynamics of the annulus dust grain medium at various strengths of magnetic field are analysed using PIV Technique. \par
In an unmagnetized plasma (B = 0 T), the confined dust particles exhibit a thermal motion around their equilibrium position. At first, the confining potential well gets modified while the magnetic field is applied. Therefore, rotational motion is not recorded at low magnetic field  (B $<$ 0.2 T). Above a magnetic field strength of B = 0.02 T, the dust grains start to rotate in anticlockwise direction. The frequency of rotational motion in the annular region is found to be nearly constant (or rigid rotation) for the conducting, as well as for the non-conducting configuration at a given magnetic field strength. The angular frequency first increases linearly with magnetic field strength up to a threshold value (B $>$ 0.6 T), and after that $\Omega$ remains nearly unchanged in the magnetic field range (0.6 T $<$ B $<$ 1 T). Above B = 1 T, a slight reduction in $\Omega$ is observed in the non-conducting configuration. The rate of change of $\Omega$ with the magnetic field strength depends on the annulus width of the dusty plasma (or the dust density) and the input rf power (or plasma density) at a given argon pressure. The saturation value of $\Omega$ shifts to a lower value as the dust width or rf power is reduced. The annulus dusty plasma of different widths exhibit an anticlockwise rigid rotational motion at different discharge conditions. At similar discharge conditions, a higher rotational frequency is expected for the conducting disk-and-ring-configuration than for the non-conducting case (observed in experiments). The uni-directional (or anticlockwise) rotation at different electrodes voltages between the electrodes and for different pressures in the presence of a magnetic field were observed. There were no discharge parameters found to observe the opposite dust flow (or shear flow) in the annular region in the presence of a magnetic field.\par 
The dust rotation in presence of a magnetic field is a result of the azimuthal ion drag force. The direction of rotation of dust grains confined by opposite radial electric fields depends on the direction of resultant radial electric field ($E_R$). The direction of rotation and the angular frequency variation with magnetic field are explained based on the magnitude of the radial potential gradient (or electric field). Since the radial potential gradient (or $E_r$) is found to be different in both disk edge region (I) and ring edge region (III), the resultant electric field ($E_R$) is expected to drive the rotational motion through $\vec{E_R} \times \vec{B}$ drift of ions. Thus, anticlockwise rotational motion of annulus dusty plasma is observed at various strengths of magnetic field. The challenges of using probe diagnostics restrict us to explain the rotational motion qualitatively. Therefore, numerical simulations are required to understand the dynamics of an annulus dusty plasma in a strong magnetic field. In future, numerical simulation of such dusty plasma experiments are planned.\\
\section{ACKNOWLEDGEMENT}
This work is supported by the Deutsche Forschungsgemeinschaft (DFG). The authors want to thank Dr. M. Kretschmer for his experimental assistance.
\bibliography{aipsamp}
\end{document}